\documentclass{ptapap}
\usepackage{color}
\newcommand{\annot}[1]{{\textbf{\textcolor{red}{#1}}}}
\usepackage{soul}

\usepackage{amsmath}
\usepackage{amssymb}
\usepackage{natbib}

\author{Chandra Shekhar Saraf}[CAMK]

\affil[CAMK]{Nicolaus Copernicus Astronomical Center, Polish Academy of Sciences, Bartycka 18, 00--716 Warsaw, Poland}

\title{Cross-Correlation Study between CMB Lensing and
Galaxy Surveys}

\begin{document}

\maketitle

\begin{abstract}

Cosmic Microwave Background (CMB) is a powerful probe to study the early universe and various cosmological models. Weak gravitational lensing affects the CMB by changing its power spectrum, but meanwhile, it also carries information
about the distribution of lensing mass and hence, the large scale structure (LSS) of the universe. When studies of the CMB is combined with the tracers of LSS, one can constrain cosmological models, models of LSS development and astrophysical parameters simultaneously. The main focus of this project is to study the cross- correlations between CMB lensing and the galaxy matter density to constrain the galaxy bias ($b$) and the amplitude scaling parameter ($A$), to test the validity of $\Lambda$CDM model. We test our approach for simulations of the Planck CMB convergence field and galaxy density field, which mimics the density field of the Herschel Extragalactic  Legacy Project (HELP). We use maximum likelihood method to constrain the parameters.

\end{abstract}

\section{Introduction}

Cosmic Microwave Background (CMB) Radiation is the thermal background of the universe resulting from the recombination in Big Bang Cosmology. It has a blackbody spectrum with mean \textit{$T_{0}$} = 2.725 K and anisotropies of the order of $\sim 10^{-5}$ K. These anisotropies tell us about the matter distribution at redshift ($z$) $\sim 1100$.\\

However, this clean picture of the universe is, to some extent, distorted by the interactions of the CMB photons with the matter distributed along its path from the surface of last scattering to us. This gravitational lensing provides us with a useful source of information on the large scale structure of the universe. Statistical signatures of lensing allow to reconstruct the gravitational potential. Gravitational lensing is an integrated measure of the matter distribution from the last scattering surface. By correlating the weak lensing field with galaxy redshift distribution, we can study the time evolution and spatial distribution of the gravitational potential. It can be used to tighten the time evolution of the dark matter density fluctuations and constraint models of structure formation and cosmological models of the Universe.\\

Here, we will focus on the last property of cross-correlation studies and estimate two parameters; galaxy bias ($b$) and amplitude of cross-spectrum ($A$) as introduced in \cite{PlanckXVII2014}. The galaxy bias is a relation between luminous tracers (galaxies, quasars, cluster of galaxies) and underlying distribution of matter. It depends on redshift and halo mass. For our study, we assume $b$ to be a constant. The amplitude of cross power spectrum is analytically expected to have a value equal to 1. Any significant departure from this value can put further questions for $\Lambda$CDM.

\section{Theory}

The lensing potential power spectrum turns out to be red, which can create some bias while estiamting the parameters. To overcome this, we introduce lensing convergence ($\kappa$) defined as

\begin{equation}
	\kappa = -\frac{\ell(\ell+1)}{2}\phi_{\ell m},
	\label{eq:lensing_potential_convergence}
\end{equation}
where $l$ corresponds to multipole of spherical harmonic expansion and $\phi$ is the lensing potential. The lensing convergence depends on the projected matter overdensity $\delta$ \citep{Bartelmann2001}

\begin{equation}
	\kappa(\hat{\textbf{n}}) = \int_{0}^{\chi_{*}} d\chi\frac{H(\chi)}{c}W^{\kappa}(\chi)\delta(\chi),
	\label{eq:convergence_density_contrast}
\end{equation}
where $\chi$ is the comoving distance, $H$ is the Hubble paramter at comoving distance $\chi$ and $\hat{\textbf{n}}$ represents the angular position on the surface of sphere. Similarly, for galaxy overdensity, we have

\begin{equation}
	g(\hat{\textbf{n}}) = \int_{0}^{\chi_{*}}d\chi\frac{H(\chi)}{c}W^{g}(\chi)\delta(\chi),
	\label{eq:galaxy_overdensity_contrast}
\end{equation}
where the lensing kernel $W^{\kappa}$ is given as

\begin{equation}
	W^{\kappa}(\chi) = \frac{3\Omega_{m}}{2c^{2}}H_{0}^{2}(1+z)\chi\frac{\chi_{*}-\chi}{\chi_{*}},
	\label{eq:lensing_kernel}
\end{equation}
and galaxy kernel is

\begin{equation}
	W^{g}(\chi) = b\frac{dN}{d\chi}+\frac{3\Omega_{m}}{2c^{2}}H_{0}^{2}(1+z)\chi\int_{\chi}^{\chi_{*}}d\chi'\frac{H(\chi')}{c}\bigg(1-\frac{\chi}{\chi'}\bigg)(\alpha(\chi')-1)\frac{dN}{d\chi'}.
	\label{eq:galaxy_kernel}
\end{equation}

In above equations, $\Omega_{m}$ and $H_{0}$ are the present day values of matter density and Hubble paramter, respectively, $c$ is the speed of light, $z$ is the redshift, $\chi_{*}$ is the comoving distance to the last scattering surface at the $z\simeq 1100$ and $\frac{dN}{d\chi}$ is the normalised galaxy redshift distribution.\\

Under Limber approximation \citep{Limber1953}, the angular power spectrum can be evaluated as

\begin{equation}
	C_{\ell}^{xy} = \int_{0}^{\chi_{*}} d\chi \frac{W^{x}W^{y}}{\chi^{2}}P(k = \frac{\ell}{\chi},z(\chi)),
	\label{eq:theoretical_power_spectrum}
\end{equation}
where $\{x,y\}=\{\kappa,g\}$ and $P(k,z)$ is the matter power spectrum calculated using CAMB (Code for Anisotropies in the Microwave Background) \citep{Lewis2000}.\\

\section{Data}

We use the lensing potential from \textit{Planck} 2018 data release described in \cite{PlanckVIII2018}. They provide lensing potential map with HEALpix \citep{Gorski2005} resolution $N_{side}=2048$, which we reduce to 512. For galaxy, we have 8 patches over the sky which is collected from HELP (Herschel Extragalactic Legacy Project) survey \citep{Raphael2019}, covering $\sim 1000$ deg$^{2}$ over sky. We have sources with photometric redshifts in the range $z=0.8$ to $\sim 7$. The sources with $z<0.8$ are discarded to reduce the contamination in evaluating the parameters. We use the posterior of these sources to compute the redshift PDF as introduced in \cite{Budavari2003}. (For more details see paper by Saraf et al. in preparation). The details for these 8 patches are summarised in Tab.~\ref{table:galaxy_patches}.

\begin{table}[h]
\label{table:galaxy_patches}
\caption{Properties of galaxy patches}

\begin{center}
\begin{tabular}{|l|cccr|}
\hline
Patch & $N_{obj}$ & $\overline{n}_{pix}$[gal pix$^{-1}$] & $\overline{n}_{str}$[gal str$^{-1}$] & median z\\
\hline
\hline
NGP & 142574 & 10.343 & 2.61$\times$10$^{6}$ & 0.8850\\
SGP & 7497737 & 359.640 & 9.00$\times$10$^{7}$ & 0.8881\\
SSDF & 3977897 & 456.547 & 1.14$\times$10$^{8}$ & 0.8772\\
LOCKMAN SWIRE & 74438 & 42.518 & 1.06$\times$10$^{7}$ & 1.3807\\
GAMA 09 & 566784 & 115.080 & 2.88$\times$10$^{7}$ & 0.9481\\
GAMA 12 & 168136 & 34.118 & 8.54$\times$10$^{6}$ & 0.9533\\
GAMA 15 & 633149 & 129.558 & 3.24$\times$10$^{7}$ & 0.9408\\
HERSCHEL STRIPE 82 & 7351928 & 332.155 & 8.31$\times$10$^{7}$ & 0.8915\\
\hline
\end{tabular}
\end{center}
\end{table}

\begin{figure}[h]
  \centering
  \begin{minipage}{0.24\textwidth}
    \includegraphics[width=\textwidth]{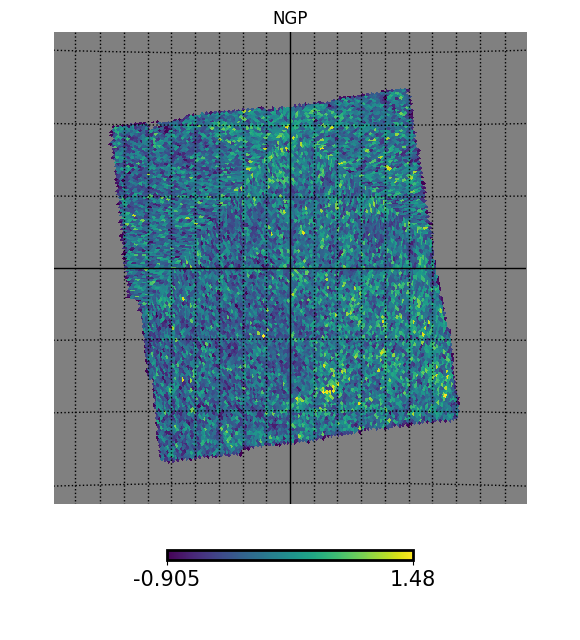}
  \end{minipage}
  \begin{minipage}{0.24\textwidth}
    \includegraphics[width=\textwidth]{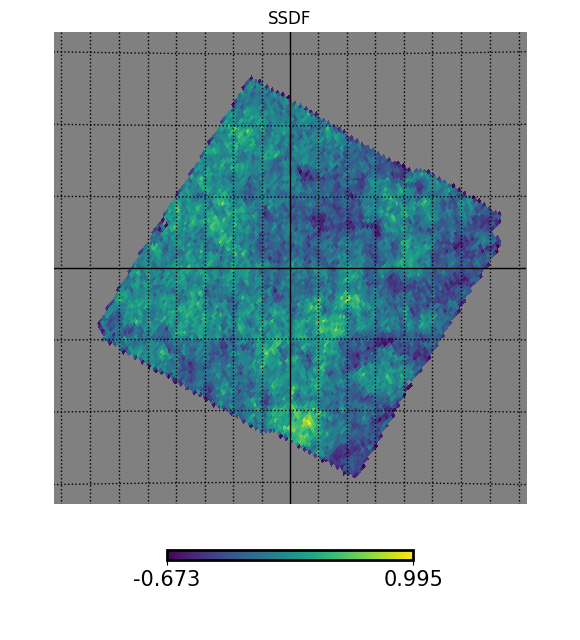}
  \end{minipage}
  \begin{minipage}{0.24\textwidth}
  	 \includegraphics[width=\textwidth]{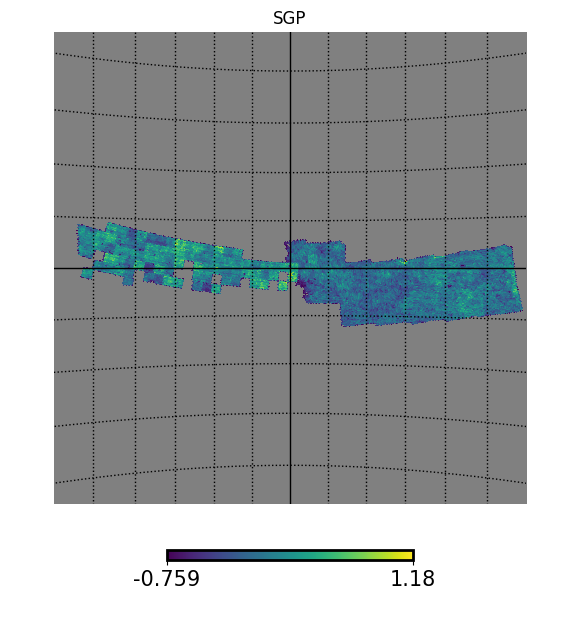}
  \end{minipage}
  \begin{minipage}{0.24\textwidth}
  	 \includegraphics[width=\textwidth]{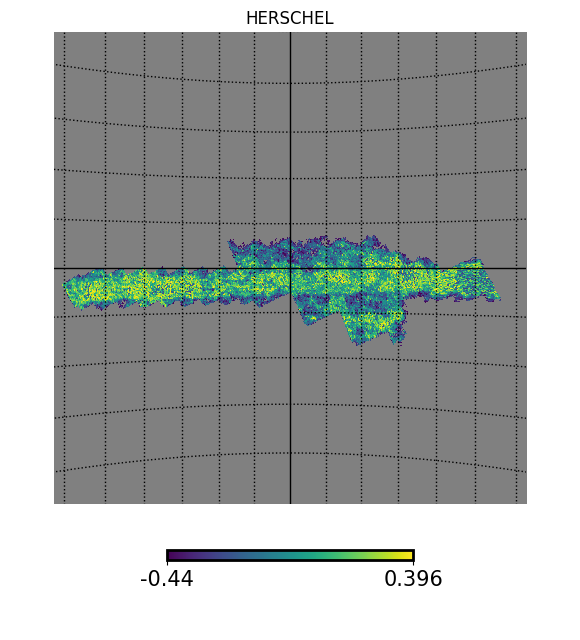}
  \end{minipage}
  \begin{minipage}{0.24\textwidth}
 	 \includegraphics[width=\textwidth]{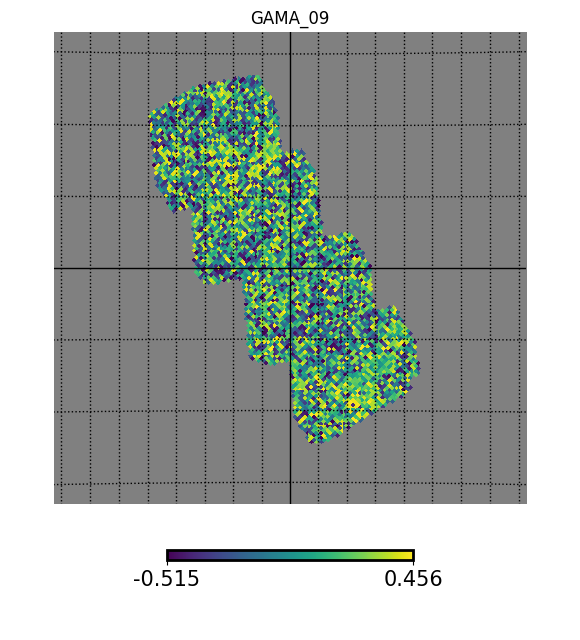}
  \end{minipage}
  \begin{minipage}{0.24\textwidth}
 	 \includegraphics[width=\textwidth]{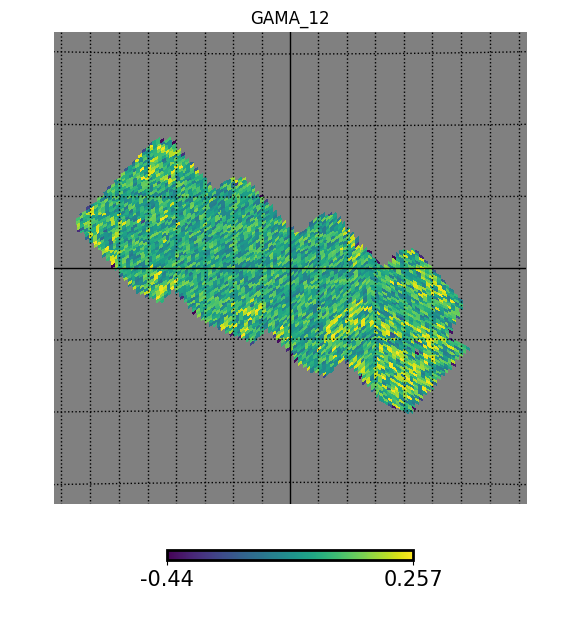}
  \end{minipage}
  \begin{minipage}{0.24\textwidth}
 	 \includegraphics[width=\textwidth]{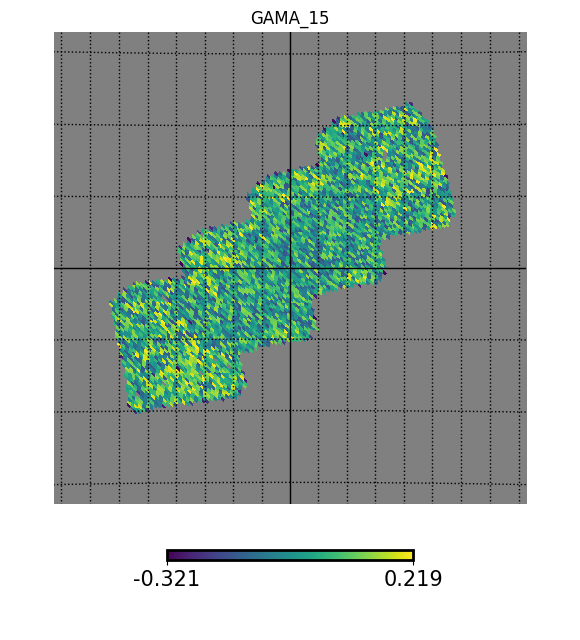}
  \end{minipage}
  \begin{minipage}{0.24\textwidth}
 	 \includegraphics[width=\textwidth]{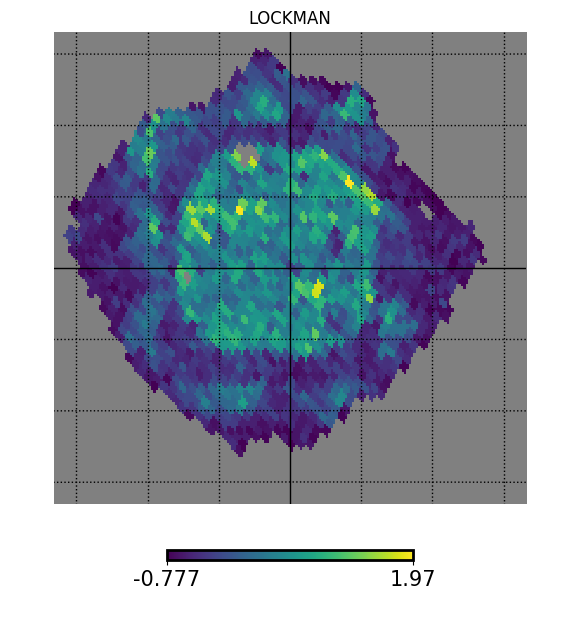}
  \end{minipage}
  \caption{Patches of galaxy on sky used in simulations. \textit{From left-}\textit{First Row:} NGP, SSDF, SGP, HERSCHEL STRIPE 82. \textit{Second Row:} GAMA 09, GAMA 12, GAMA 15, LOCKMAN SWIRE.}
  \label{fig:gal_patches}
\end{figure}

In Fig \ref{fig:gal_patches}, we show the position of patches on sky with galactic coordinates.


\section{Methodology}

For each patch, we simulate 100 maps for both \textit{Planck} lensing convergence and galaxy overdensity, with $b=2$ and $A=1$, and introducing a known degree of correlation in the theoretical power spectra given in Eq. \ref{eq:theoretical_power_spectrum}. The correlation is inserted through complex random drawn from gaussian distribution \citep{Kamionkowski1997}

\begin{equation}
\begin{aligned}
\kappa_{\ell m} &= \alpha_{1}(C_{\ell}^{\kappa\kappa})^{1/2}; \\
g_{\ell m} &= \alpha_{1}\frac{C_{\ell}^{\kappa g}}{(C_{\ell}^{\kappa\kappa})^{1/2}}+\alpha_{2}\bigg[C_{\ell}^{gg}-\frac{(C_{\ell}^{\kappa g})^{2}}{C_{\ell}^{\kappa\kappa}}\bigg]^{1/2}\annot{.}
\end{aligned}
\label{eq:simuleqn}
\end{equation}

For each $\ell$ and $m > 0$, $\alpha_{1}$ and $\alpha_{2}$ are two complex random numbers drawn from a Gaussian distribution with unit variance, and for $m=0$, $\alpha_{1}$ and $\alpha_{2}$ are normally distributed real random numbers. Further we add noise to these maps, based on the lensing convergence noise provided in \textit{Planck} 2018 data release and mean number of sources per solid angle (as shown in Tab.~\ref{table:galaxy_patches}) for galaxy maps. (For details on the procedure, see \cite{Bianchini2015}).\\

We use the MASTER algorithm \citep{Hivon2002} to estimate the full sky power spectrum for convergence, galaxy overdensity and their cross-sepctra. The parameters, galaxy bias ($b$) and amplitude of cross-spectrum ($A$) is then computed using EMCEE, which is affine invariant Markov Chain Monte Carlo Ensemble sampler (\cite{EMCEE2013}). These parameters are estimated using Maximum Likelihood Estimation (MLE) approach. For this, we use three likelihood functions, which are assumed to be Gaussian. Using multiple likelihood functions has two-fold advantage; first, it breaks the degeneracy that exists between $b$ and $A$ and second, it also checks for systematics in parameter estimation. For details on this expression and likelihood functions, see Saraf et al. (in preparation).

\begin{equation}
\begin{split}
	Cov_{bb'}^{AB,CD} = &(M^{AB^{-1}}_{bb_{1}}P_{b_{1}\ell})\bigg[\frac{M^{AC,BD}_{\ell\ell '}}{(2\ell '+1)}\sqrt{C_{\ell}^{AC}C_{\ell '}^{AC}C_{\ell}^{BD}C_{\ell '}^{BD}}\\
	&+\frac{M^{AD,BC}_{\ell\ell '}}{(2\ell '+1)}\sqrt{C_{\ell}^{AD}C_{\ell '}^{AD}C_{\ell}^{BC}C_{\ell '}^{BC}}\bigg] (M^{CD^{-1}}_{bb_{2}}P_{b_{2}\ell '})^{T},
\end{split}
	\label{eq:error_covariance}
\end{equation}
where $M_{\ell\ell '}$ and \annot{ }$M_{bb '}$ are the coupling kernel and binned coupling kernel, respectively obtained from MASTER algorithm, $P_{b\ell}$ is the binning operator, $T\equiv$ Transpose operator and $\{A,B,C,D\}=\{\kappa,g\}$. For details on this expression of covariance, see \cite{Tristram2005}.

\section{Results and Conclusions}

\begin{figure*}[t]
 \centering
 \begin{minipage}{0.48\textwidth}
 	 \includegraphics[width=\textwidth]{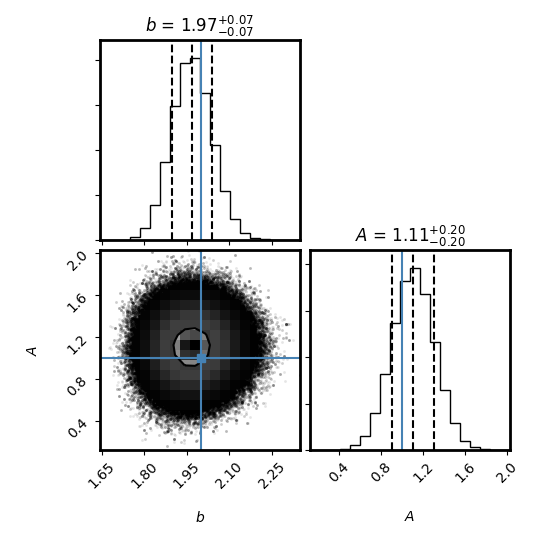}
 \end{minipage}
 \begin{minipage}{0.48\textwidth}
 	 \includegraphics[width=\textwidth]{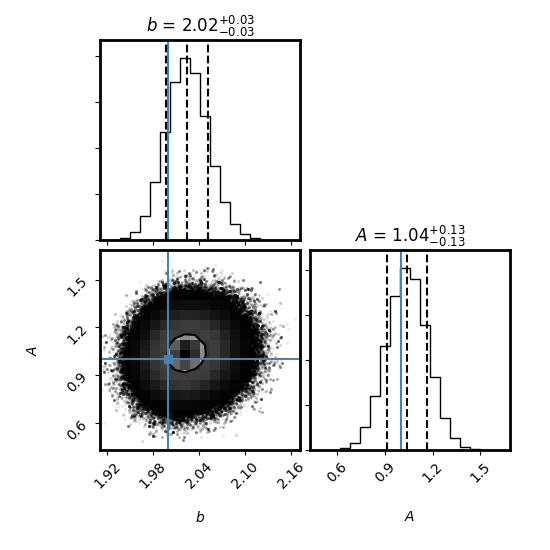}
 \end{minipage}
 \begin{minipage}{0.48\textwidth}
 	 \includegraphics[width=\textwidth]{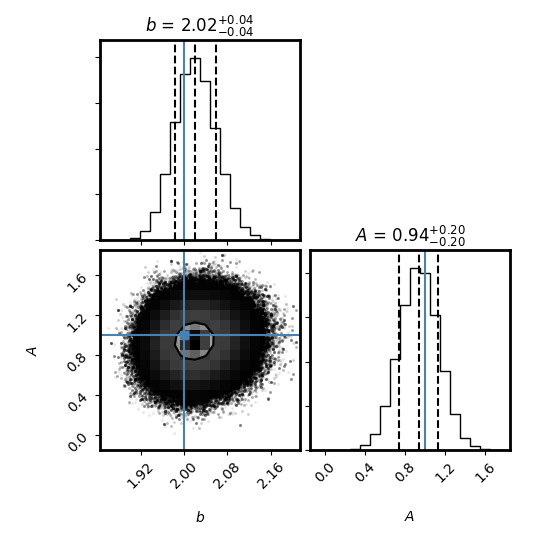}
 \end{minipage}
 \begin{minipage}{0.48\textwidth}
 	 \includegraphics[width=\textwidth]{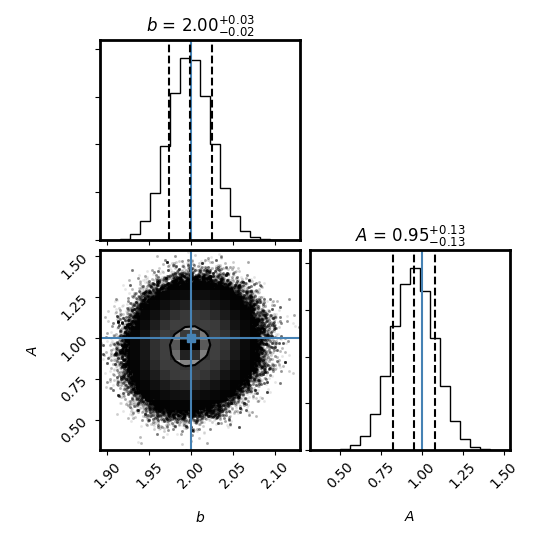}
 \end{minipage}
 \caption{Results of parameter estimation using MLE. \textit{From top left}: NGP, SGP, SSDF and HERSCHEL STRIPE 82.}
 \label{fig:joint_likeli_simulations}
\end{figure*}

\begin{figure*}
\centering
 \begin{minipage}{0.32\textwidth}
 	 \includegraphics[width=\textwidth]{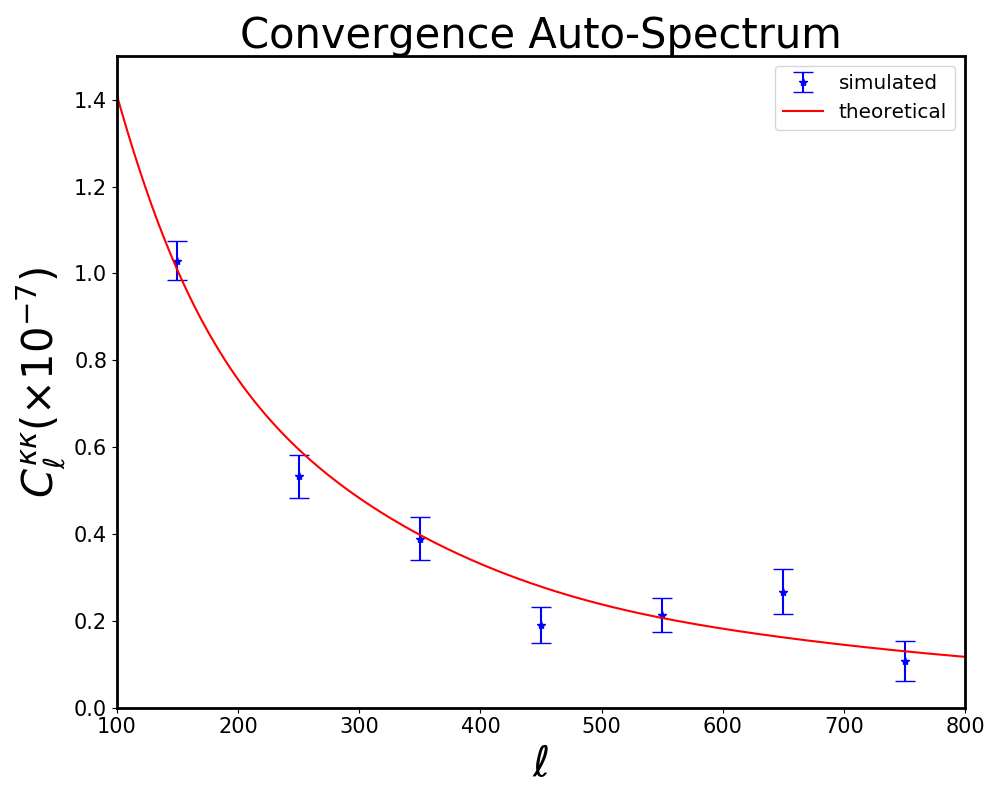}
 \end{minipage}
 \begin{minipage}{0.32\textwidth}
 	 \includegraphics[width=\textwidth]{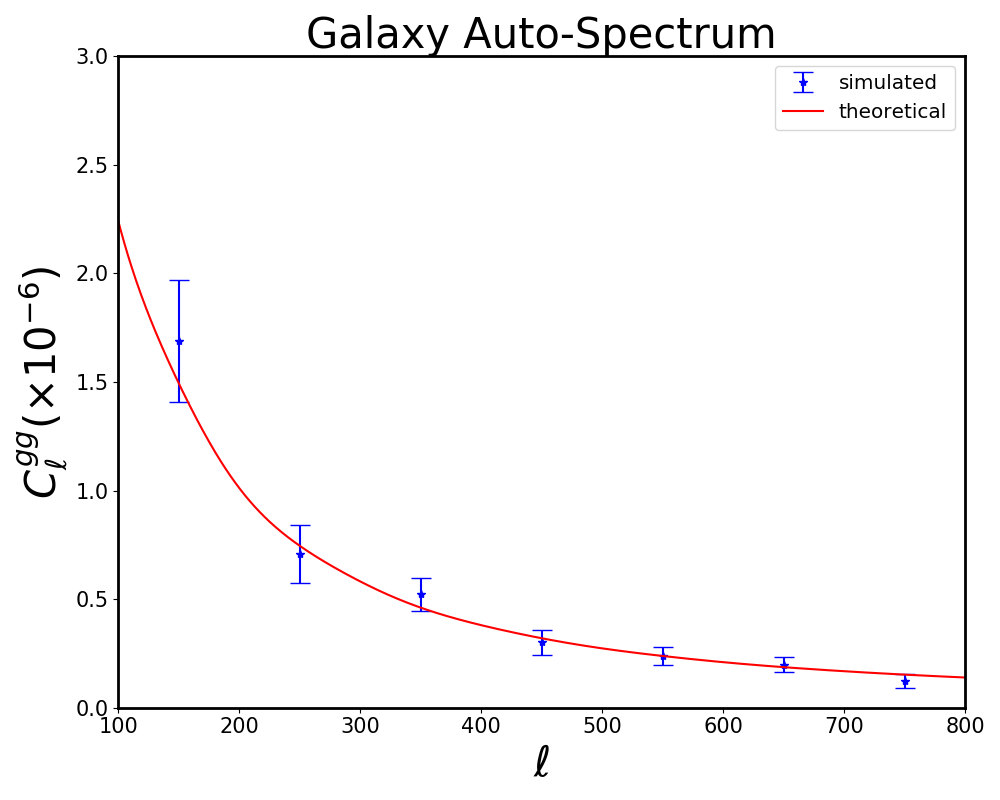}
 \end{minipage}
 \begin{minipage}{0.32\textwidth}
 	 \includegraphics[width=\textwidth]{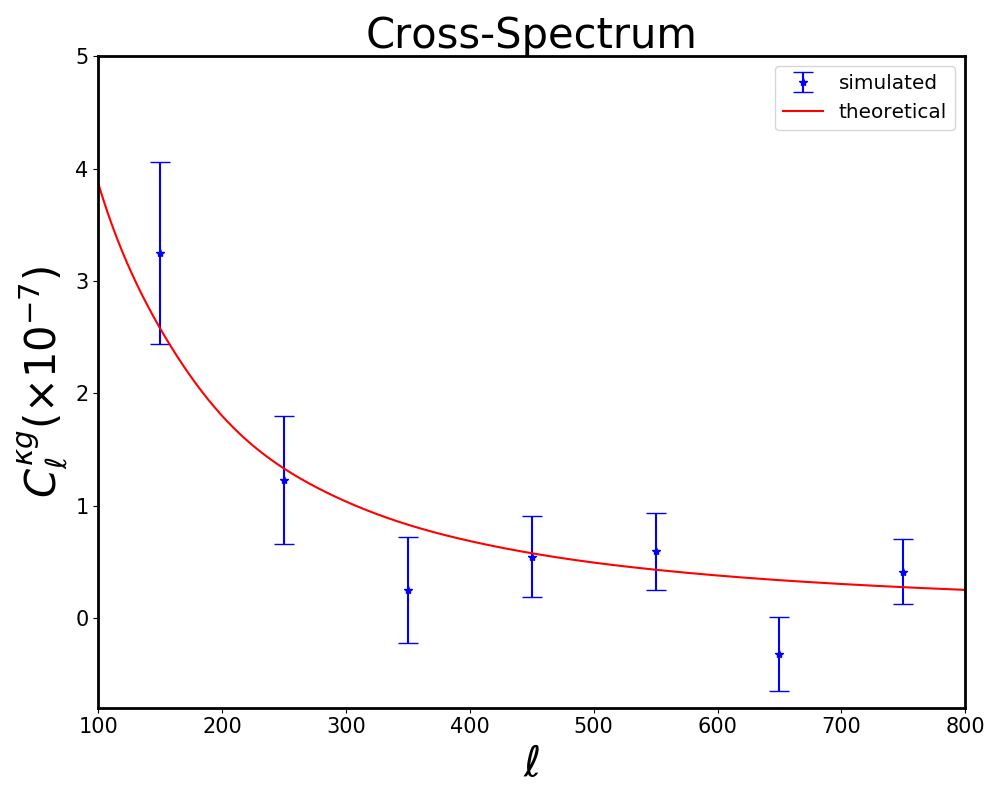}
 \end{minipage}
 \caption{Power spectra from one realisation recovered from simulation with errors. \textit{From left}: convergence-, galaxy- and cross-spectra for NGP. The errors are the diagonal of covariance matrix obtained from simulations. }
 \label{fig:power_spectra_simulations}
\end{figure*}

We see that the power spectra are recovered very well from our algorithms without any significant indication of systematic bias in estimation. With this power spectra, we estimate the parameters using MLE. The results for patches NGP, SGP, SSDF and HERSCHEL STRIPE 82 are as shown in Fig.~\ref{fig:joint_likeli_simulations}.
Fig.~\ref{fig:power_spectra_simulations} clearly validates our numerical approach in estimating the parameters. For all the patches, we recovered the parameters within $1\sigma$ error range. With this we can use our procedure to analyse real data.\\

\section{Summary}

We here presented the results of our simulation setup, prepared to analyse and study the properties of largescale structure development and $\Lambda$CDM from cross correlation between \textit{Planck} lensing convergence field and galaxy survey from HELP surveys. So far, it is the largest collection of sources with photometric redshifts. Hence, it provides an unpreceeded evalutaion of parameters from the above-mentioned datasets. The analysis of the real data and the discussion henceforth, will be presented in Saraf et al. 2020 (in preparation). We put further constraints on data to obtain a homogeneous patch, which is very important for an unbiased estimation of $b$ and $A$.\\

We thank Kenneth Duncan, Raphael Shirley and Katarzyna Ma{\l}ek for their help in understanding the galaxy fields. C.S.S. would like to thank Pawel Bielewicz for counteless discussions and ideas put forth in developing the simulation setup.\\


\bibliographystyle{ptapap}
\bibliography{saraf}

\begin{thebibliography}{13}
\providecommand{\natexlab}[1]{#1}
\providecommand{\url}[1]{\texttt{#1}}
\providecommand{\urlprefix}{URL }
\providecommand{\eprint}[2][]{\url{#2}}

\bibitem[{{Bartelmann} \& {Schneider}(2001)}]{Bartelmann2001}
{Bartelmann}, M., {Schneider}, P., \emph{\physrep} \textbf{340}, 4-5, 291
  (2001)

\bibitem[{{Bianchini} et~al.(2015)}]{Bianchini2015}
{Bianchini}, F., et~al., \emph{\apj} \textbf{802}, 1, 64 (2015)

\bibitem[{{Budav{\'a}ri} et~al.(2003)}]{Budavari2003}
{Budav{\'a}ri}, T., et~al., \emph{\apj} \textbf{595}, 1, 59 (2003)

\bibitem[{{Foreman-Mackey} et~al.(2013){Foreman-Mackey}, {Hogg}, {Lang}, \&
  {Goodman}}]{EMCEE2013}
{Foreman-Mackey}, D., {Hogg}, D.~W., {Lang}, D., {Goodman}, J., \emph{\pasp}
  \textbf{125}, 925, 306 (2013)

\bibitem[{{G{\'o}rski} et~al.(2005)}]{Gorski2005}
{G{\'o}rski}, K.~M., et~al., \emph{\apj} \textbf{622}, 2, 759 (2005)

\bibitem[{{Hivon} et~al.(2002)}]{Hivon2002}
{Hivon}, E., et~al., \emph{\apj} \textbf{567}, 1, 2 (2002)

\bibitem[{{Kamionkowski} et~al.(1997){Kamionkowski}, {Kosowsky}, \&
  {Stebbins}}]{Kamionkowski1997}
{Kamionkowski}, M., {Kosowsky}, A., {Stebbins}, A., \emph{\prd} \textbf{55},
  12, 7368 (1997)

\bibitem[{{Lewis} et~al.(2000){Lewis}, {Challinor}, \& {Lasenby}}]{Lewis2000}
{Lewis}, A., {Challinor}, A., {Lasenby}, A., \emph{\apj} \textbf{538}, 2, 473
  (2000)

\bibitem[{{Limber}(1953)}]{Limber1953}
{Limber}, D.~N., \emph{\apj} \textbf{117}, 134 (1953)

\bibitem[{{Planck Collaboration} et~al.(2014)}]{PlanckXVII2014}
{Planck Collaboration}, et~al., \emph{\aap} \textbf{571}, A17 (2014)

\bibitem[{{Planck Collaboration} et~al.(2018)}]{PlanckVIII2018}
{Planck Collaboration}, et~al., \emph{arXiv e-prints} arXiv:1807.06210 (2018)

\bibitem[{{Shirley} et~al.(2019)}]{Raphael2019}
{Shirley}, R., et~al., \emph{\mnras} \textbf{490}, 1, 634 (2019)

\bibitem[{{Tristram} et~al.(2005){Tristram}, {Mac{\'\i}as-P{\'e}rez},
  {Renault}, \& {Santos}}]{Tristram2005}
{Tristram}, M., {Mac{\'\i}as-P{\'e}rez}, J.~F., {Renault}, C., {Santos}, D.,
  \emph{\mnras} \textbf{358}, 3, 833 (2005)

\end{thebibliography}

\end{document}